\documentclass[11pt,fleqn,twoside]{article}
\usepackage{amsfonts,latexsym}
\makeatletter
\newcommand{\prava}{\footnotesize\it
\begin{flushright}
\begin{minipage}{18cm}
Copyright \copyright 1998 by M. Senthil Velan and M. Lakshmanan
\end{minipage}
\end{flushright}}

\newcommand{\name}[1]{\begin{flushleft}
                       \LARGE \bf #1
                       \end{flushleft}\vspace{-3mm}}

\newcommand{\Author}[1]{\begin{flushleft}
                       \it #1 \end{flushleft}}

\newcommand{\Adress}[1]{\begin{flushleft}
                       \it #1 \end{flushleft}}

\newcommand{\Date}[1]{\begin{flushleft}
                      \small  \it #1 \end{flushleft}}

\newcommand{\ehkol}{Author \ name}
\newcommand{\ohkol}{Article \ name}
\renewcommand{\@evenhead}{
\hspace*{-3pt}\raisebox{-15pt}[\headheight][0pt]{\vbox{\hbox to \textwidth
{\thepage \hfil \ehkol}\vskip4pt \hrule}}}
\renewcommand{\@oddhead}{
\hspace*{-3pt}\raisebox{-15pt}[\headheight][0pt]{\vbox{\hbox to \textwidth
{\ohkol \hfil \thepage}\vskip4pt\hrule}}}
\renewcommand{\@evenfoot}{}
\renewcommand{\@oddfoot}{}

\newcommand{\be}{\begin{equation}}
\newcommand{\ee}{\end{equation}}
\newcommand{\ba}{\hspace*{-5pt}\begin{array}}
\newcommand{\ea}{\end{array}}

\newcommand{\ds}{\displaystyle}
\makeatother

\begin{document}

\setcounter{page}{190}

\thispagestyle{empty}

\renewcommand{\ehkol}{M. Senthil Velan and M. Lakshmanan}
\renewcommand{\ohkol}{Lie Symmetries, Kac-Moody-Virasoro Algebras and
Integrability}

\begin{flushleft}
\footnotesize \sf
Journal of Nonlinear Mathematical Physics \qquad 1998, V.5, N~2,
\pageref{lakshmanan-fp}--\pageref{lakshmanan-lp}. \hfill {\sc Article}
\end{flushleft}

\vspace{-5mm}

\renewcommand{\footnoterule}{}
{\renewcommand{\thefootnote}{}
 \footnote{\prava}}

\name{Lie Symmetries, Kac-Moody-Virasoro Algebras and Integrability
of Certain (2+1)-Dimensional Nonlinear Evolution Equations}
\label{lakshmanan-fp}

\Author{M. SENTHIL VELAN and M. LAKSHMANAN~$^*$}

\Adress{Centre for Nonlinear Dynamics, Department of Physics,\\
Bharathidasan University, Tiruchirapalli 620 024, India\\[1mm]
$^*$~E-mail: lakshman@bdu.ernet.in}

\Date{Received March 06, 1998}

\begin{abstract}
\noindent
In this paper we study Lie symmetries, Kac-Moody-Virasoro algebras,
similarity reductions and particular solutions of two dif\/ferent recently
introduced (2+1)-di\-men\-sio\-nal nonlinear evolution equations, namely (i)
(2+1)-dimensional breaking soliton equation and (ii) (2+1)-di\-men\-sio\-nal
nonlinear
Schr\"{o}dinger type equation introduced by Zakharov and studied later
by Strachan.
Interestingly our studies show that not all integrable higher dimensional
systems admit Kac-Moody-Virasoro type sub-algebras.  Particularly
the two integrable systems
mentioned above do not admit Virasoro type subalgebras, eventhough the other
integrable higher dimensional systems do admit such algebras
which we have also reviewed in
the Appendix.  Further, we bring out physically interesting solutions for
special choices of the symmetry parameters in both the systems.
\end{abstract}

\section{Introduction}

In recent years important progress has been made in the understanding of
(2+1)-di\-men\-sio\-nal nonlinear evolution equations (NLEEs) and their methods of
solution [1, 2]. In this direction it is well realized that the Lie group
method [3--8], originally introduced by Sophus Lie, can play a crucial role,
since in most of the problems it not only explores the intrinsic geometric
properties but also brings out interesting physical solutions in a
straightforward manner. Eventhough the last decade has witnessed a veritable
explosion on the applications of this method to explore the invariance and
integrability properties of a large class of problems in (1+1)-dimensions
[6, 9--11] very few systems in higher dimensions have been
explored in this way.

Recently the invariance properties of some of the physically important
integrable NLEEs$\!$ in (2+1)-dimensions, such as the Kadomtsev-Petvishilli
equation~[12], the Davey-Ste\-war\-tson equation~[13], the three wave
interaction problem~[14], cylindrical Kadomtsev-Petviashvili equations~[15]
and stimulated Raman scattering equation~[16] have been studied through
Lie symmetry analysis and
it has been shown that all these equations admit inf\/inite dimensional Lie
point symmetry groups with a specif\/ic Kac-Moody-Virasoro structure.
Further, the present authors have also carried out a detailed study on the
invariance properties of certain higher-dimensional nonlinear evolution
equations, namely, (i) Nizhnik-Novikov-Veselov equation, (ii) breaking
soliton equation, (iii) nonlinear Schr\"{o}\-din\-ger type equation
studied by Fokas recently, (iv) sine-Gordon equation and (v)
(2+1)-di\-men\-sio\-nal
long dispersive wave equation introduced by Chakravarthy, Kent
and Newman and explored
possible similarity reductions and Kac-Moody-Virasoro algebras and also
particular solutions associated with them~[17, 18]. While all the
above mentioned equations except the breaking soliton equation
admit Virasoro type subalgebras the later one does not admit such subalgebras.

In contradistinction to the above integrable NLEEs, the (2+1)-dimensional
nonintegrable partial dif\/ferential equations (PDEs)  do not admit Virasoro
type subalgebras. Typical examples are Infeld-Rowlands equation~[19], and a
nonintegrable dispersive long-wave equation~[20]. Thus it is commonly
believed in the current literature that all integrable higher dimensional
NLEEs will admit Virasoro type subalgebras while nonintegrable equations do
not. However in this paper we wish to point out that not all the integrable
(2+1)-dimensional NLEEs admit Virasoro type algebras. For example, we have
carried out a detailed investigation on the invariance properties of two
dif\/ferent classes of NLEEs, namely, (i) breaking
soliton equation~[21] which is an asymmetric generalization of the
Korteweg de Vries~(KdV)
equation in (2+1)-dimensions and (ii) (2+1)-dimensional nonlinear
Schr\"{o}dinger equation (NLS) studied recently by Strachan~[22]. We
have found that both the systems do not admit Kac-Moody-Virasoro type
subalgebras. However it has been shown in the literature~[21--24] that
both the systems are in fact integrable.

Another interesting feature of our study is that both the equations,
namely, the breaking soliton equation and the Zakharov-Strachan
equation, eventhough belong to two
dif\/ferent categories, namely KdV and NLS type respectively, admit a
specif\/ic type of symmetries (see eqs.(\ref{lakshmanan:2.4}) and (\ref{lakshmanan:3.6}) given below).
For example, both the equations allow the inf\/initesimals upto
quadratic power in $t$ explicitly.  Also they do not admit any
arbitrary function in the inf\/initesimal variations in $t$ which
inturn leads to the absence of Kac-Moody-Virasoro type subalgebras in
both the systems.  However, the other integrable nonlinear evolution
equations, given in the Appendix,
admit arbitrary functions in the inf\/initesimal transformation of $t$.

Further, we also bring out the unexplored invariance properties of the above
two NLEEs through Lie group method. First we obtain the appropriate point
transformation groups and generators, which are inf\/inite in both the
systems, which leave the above two nonlinear systems invariant. By solving
the characteristic equation associated with the inf\/initesimals we obtain the
similarity variables interms of which the original system with three
independent variables reduce to a PDE with two independent variables. For
the latter, again another set of similarity variables are found interms of
which the PDE reduces to an ordinary dif\/ferential equation (ODE). We have
used the symbolic manipulation program LIE~[25] to f\/ind out the Lie
symmetries.

The plan of the paper is as follows. In Appendix A, we briefly summarize
the Lie symmetries and Kac-Moody-Virasoro algebras of certain NLEEs
discussed in the literature. In Sec.~2, we present the symmetry algebra of
the breaking soliton equation and its similarity reductions. In Sec.~3, we
report the Lie symmetries and Kac-Moody-Virasoro algebras of the nonlinear
Schr\"{o}dinger equation studied by Strachan. Further we have also
explored the
possible similarity reductions and particular solutions. In Sec.~4 we present
our conclusions.

\renewcommand{\thesection}{\arabic{section}}

\setcounter{equation}{0}
\renewcommand{\theequation}{\arabic{section}.\arabic{equation}}

\section{Lie symmetries and Kac-Moody-Virasoro algebras of the breaking
soliton equation}

An asymmetric generalization of the KdV equation in (2+1)-dimensions~[21] is
\be \label{lakshmanan:2.1}
\ba{l}
u_t+Bu_{xxy}+4Buv_x+2Bvu_x = 0,\\
u_y = v_x,
\ea
\ee
which describes the interaction of a Riemann wave propagating along the $y$
axis with a long wave propagating along the $x$ axis.
Eq.(\ref{lakshmanan:2.1}) can also be
written as the single fourth order nonlinear PDE of the form,
\[
\rho_{xt}+\rho_{xxxy}+4B\rho_x \rho_{xy}+2B\rho_y \rho_{xx} = 0 ,
\]
by introducing the transformation, $u=\rho_x$, $v=\rho_y$. Eq.(\ref{lakshmanan:2.1})
admits Lax representation~[21]. One can also easily verify that eq.(\ref{lakshmanan:2.1})
admits Painlev\'e property. Special
features of eq.(\ref{lakshmanan:2.1}) have also been studied extensively in ref.~[21].

To study the invariance properties we have considered the equation of
the form (\ref{lakshmanan:2.1}). The invariance of eq.(\ref{lakshmanan:2.1}) under the inf\/initesimal point
transformations
\[
\ba{l}
x\longrightarrow X = x+\varepsilon\xi_1(t,x,y,u,v),\\[1mm]
y\longrightarrow Y = y+\varepsilon\xi_2(t,x,y,u,v),\\[1mm]
t\longrightarrow T = t+\varepsilon\xi_3(t,x,y,u,v),\\[1mm]
u\longrightarrow U = u+\varepsilon\phi_1(t,x,y,u,v),\\[1mm]
v\longrightarrow V = v+\varepsilon\phi_2(t,x,y,u,v),  \quad \varepsilon \ll 1
\ea
\]
leads to the expressions for the inf\/initesimals
\be \label{lakshmanan:2.4}
\ba{l}
\ds \xi_1 = -\frac{c_1}{3}xt-\frac{c_2}{2}x+f(t),\\[2mm]
\ds \xi_2 =
-\frac{2c_1}{3}yt+\left(\frac{c_2}{2}-c_4\right)y-4Bc_3t+c_5,\\[2mm]
\ds \xi_3 = -\frac{2c_1}{3}t^2-\left(\frac{c_2}{2}+c_4\right)t-c_6,\\[2mm]
\ds \phi_1 = \frac{2c_1}{3}ut+c_2u-\frac{c_1}{6B}y-c_3,\\[2mm]
\ds \phi_2 = c_1vt+c_4v-\frac{c_1}{6B}x+\frac{\dot{f}(t)}{2B},
\ea
\ee
where $ c_1, c_2, c_3, c_4, c_5, c_6$ are arbitrary constants and $f(t)$
is an arbitrary function of $t $ and dot denotes
dif\/ferentiation with respect to $t$.

\subsection{Lie algebra of symmetry vector f\/ields}

The presence of the arbitrary function $f$ of $t $ leads to an inf\/inite
dimensional Lie algebra of symmetries. We can write a general element of
this Lie algebra as
\[
V=V_1(f)+V_2+V_3+V_4+V_5+V_6,
\]
where
\[
\ba{l}
\ds V_1(f) = f(t)\frac{\partial}{\partial x} +
\frac{\dot {f}(t)}{2B}\frac{\partial}{\partial v} ,\\[4mm]
\ds V_2 = -\frac{1}{3}xt \frac{\partial}{\partial x}- \frac{2}{3}yt \frac{%
\partial}{\partial y}- \frac{2}{3}t^2 \frac{\partial}{\partial t}+
\left(\frac{2}{3}ut-\frac{y}{6B}\right)
\frac{\partial}{\partial u}+
\left(vt-\frac{x}{6B}\right)\frac{\partial}{\partial v},
\\[4mm]
\ds V_3 = -\frac{1}{2}x \frac{\partial}{\partial x}+ \frac{1}{2}y \frac{\partial%
}{\partial y}- \frac{1}{2}t \frac{\partial}{\partial t}+ u \frac{\partial}{%
\partial u}, \quad
 V_4 = -4Bt \frac{\partial}{\partial y}-\frac{\partial}{\partial u},
\\[4mm]
\ds V_5 = -y\frac{\partial}{\partial y}-t\frac{\partial}{\partial t} +v\frac{
\partial}{\partial v},
\quad V_6 = \frac{\partial}{\partial y},\quad
V_7 = \frac{\partial}{\partial t}.
\ea
\]
The associated Lie algebra between these vector f\/ields becomes
\[
\ba{l}
\ds \left[V_1, V_2\right] = V_1\left(-\frac{1}{3}tf+ \frac{2}{3}t^2 \dot {f}
\right),\quad \left[V_1, V_3\right] = V_1\left(\frac{f}{2}
+\frac{1}{2}t \dot {f}\right), \quad
\left[V_1, V_4\right] = 0, \\[2mm]
\ds  \left[V_1, V_5\right] = V_1(t \dot {f}),
\quad  \left[V_1, V_6\right] = 0,\quad
 \left[V_1, V_7\right] = -V_1(\dot {f}),\quad
\left[V_2, V_3\right] = \frac{V_2}{2},\\[2mm]
\ds  \left[V_2, V_4\right] = 0,\quad
\left[V_2, V_5\right] = V_2,
\quad \left[V_2, V_6\right] = -\frac{1}{6B}V_4, \quad
 \left[V_2, V_7\right] = -\frac{2}{3}V_3-V_5,
\\[3mm]
\ds  \left[V_3, V_4\right] = -V_4,\quad
\left[V_3, V_5\right] = 0,\quad \left[V_3, V_6\right] = -\frac{1}{2}V_6,
\quad
 \left[V_3, V_7\right] = \frac{1}{2}V_7, \\[2mm]
\ds \left[V_4, V_5\right] = 0,
\quad
  \left[V_4, V_6\right] = 0,\quad  \left[V_4,
V_7\right] = 4BV_6, \quad  \left[V_5, V_6\right] = V_6,\\[2mm]
\ds  \left[V_5, V_7\right] = V_7,\quad
\left[V_6, V_7\right] = 0,
\ea
\]
whereas the commutation relation between $V_1(f_1), V_1(f_2)$ turns out to
be
\[
\left[V_1(f_1), V_1(f_2)\right] = 0,
\]
which is not of   {\it Virasoro type} which typically exists in
most of the
integrable systems mentioned in the introduction and also pointed out in the
Appendix.

\subsection{Similarity variables and similarity reductions}

The similarity variables associated with the inf\/initesimal symmetries (\ref{lakshmanan:2.4})
can be obtained by solving the associated invariant surface condition or the
related characteristic equation. The latter reads
\be \label{lakshmanan:2.9}
\ba{l}
\ds \frac{dx}{ -\frac{c_1}{3}xt-\frac{c_2}{2}x+f(t)} =
\frac{dy}{ -\frac{2}{3}
c_1yt+\left(\frac{c_2}{2}-c_4\right)y-4Bc_3t+c_5}\\[4mm]
\ds \qquad\qquad =\frac{dt}{ -\frac{2}{3}c_1t^2-\left(\frac{
c_2}{2}+c_4\right)t-c_6} =\frac{du}
{ \frac{2}{3}c_1ut+c_2u-\frac{1}{6B}c_1y-c_3}
 \\[4mm]
 \ds \qquad\qquad =\frac{dv}{-\frac{c_1}{6B}x+\frac{1}{2B}\dot{f}(t)+c_1vt+c_4v}.
 \ea
 \ee
Integrating eq.(\ref{lakshmanan:2.9}) with the condition $c_1 \neq 0$ we get the following
similarity variables:
\[
\ds \tau_1 = \frac{x(t+k_1+k_2)^{n-(1/4)}}{(t+k_1-k_2)^{n+(1/4)}} +
\frac{3}{2c_1} \int_{0}^{t} \frac{f(t^{\prime})
(t^{\prime}+k_1+k_2)^{n-(5/4)}} {(t^{
\prime}+k_1-k_2)^{n+(5/4)}}dt^{\prime},
\]
\[
\ds \tau_2 = \frac{y(t+k_1+k_2)^{-(2n+(1/2))}}
{(t+k_1-k_2)^{-2n+(1/2)}}- \frac{3}{2c_1}
\int_{0} ^{t} \frac{(4Bc_3t^{\prime}-c_5)(t^{\prime}+k_1+k_2)^{-(2n+(3/2))}} {%
(t^{\prime}+k_1-k_2)^{-2n+(3/2)}}dt^{\prime},
\]
\[
\ba{l}
\ds
F = \frac{u(t+k_1+k_2)^{2n+(1/2)}}{(t+k_1-k_2)^{2n-(1/2)}}\\[4mm]
\ds \qquad - \frac{3}{8Bc_1}\int_{0} ^{t}
\left[ \int_{0} ^{t^{\prime}} \frac{(4Bc_3 t^{\prime\prime}-c_5)
(t^{\prime\prime}+k_1+k_2)^{-(2n+(3/2))}} {(t^{\prime\prime}+k_1-k_2)^{-2n+(3/2)}}%
dt^{^{\prime\prime}}\right] dt^{\prime}\\[4mm]
\ds \qquad  -\frac{\tau_2 t}{4B} - \frac{3c_3}{2c_1}\int_{0} ^{t}
\frac{(t^{\prime}+k_1-k_2)^{2n-(1/2)}}
{(t^{\prime}+k_1+k_2)^{2n+(1/2)}}dt^{\prime},
\ea
\]
\[
\ba{l}
\ds G = \frac{v(t+k_1+k_2)^{n+(3/4)}}{(t+k_1-k_2)^{n-(3/4)}}+
\frac{3}{8Bc_1}\int_{0} ^{t}
\left[ \int_{0} ^{t^{\prime}} \frac{f(t^{\prime\prime})(t^{\prime
\prime}+k_1+k_2)^{n-(5/4)}} {(t^{\prime\prime}+k_1-k_2)^{n+(5/4)}}
dt^{^{\prime\prime}} \right] dt^{\prime}
\\[4mm]
\ds \qquad -\frac{\tau_1 t}{4B}+\frac{3}{4Bc_1}
\int_{0} ^{t} \frac{\dot {f}
(t^{\prime})(t^{\prime}+k_1-k_2)^{-(n+(1/4))}}
 {(t^{\prime}+k_1+k_2)^{-n+(1/4)}}dt^{\prime},
 \ea
\]
where $F$ and  $G$ are functions of $\tau_1$ and $\tau_2$ and
\[
\ba{l}
\ds k_1 = (3c_2+6c_4)/8c_1, \qquad
 k_2 = \sqrt{(3c_2+6c_4)^2-96c_1c_6}/8c_1, \\[3mm]
\ds n = (9c_2-6c_4)/4\sqrt{(3c_2+6c_4)^2-96c_1c_6}.
\ea
\]
Under the above similarity transformations, eq.(\ref{lakshmanan:2.1}) gets reduced to a
system of PDEs in two independent variables $\tau_1$ and $\tau_2$:
\be \label{lakshmanan:2.11}
\ba{l}
\ds BF_{{\tau_1}{\tau_1}{\tau_2}}+4BFG_{\tau_1}+2BGF_{\tau_1}
+\frac{(3c_2 -6c_4)} {4c_1}\tau_2F_{\tau_2}
\\[2mm]
\ds \qquad \qquad -\frac{3c_2}{4c_1}\tau_1 F_{\tau_1}
 - \frac{3c_2}{2c_1}F + \frac{3c_6}{8Bc_1}\tau_2= 0,
\qquad
F_{\tau_2} = G_{\tau_1}.
\ea
\ee
Since the original (2+1)-dimensional PDE (\ref{lakshmanan:2.1}) satisf\/ies the Painlev\'{e}
property for a general manifold, the (1+1)-dimensional similarity reduced
PDE (\ref{lakshmanan:3.11}) will also naturally satisfy the P-property and so is a candidate
for a completely integrable system in (1+1)-dimensions.

\subsection{Subcases}
\label{sec:2.3}
In addition to the above general similarity reduction one can also look into
the subcases by assuming one or more of the vector f\/ields to be zero. We
have considered all the subcases and in the following we report only the
distinct nontrivial cases.

\noindent
{\bf Case 1:} $c_1 = 0$. The similarity variables are
\[
\ds \tau_1 = \frac{x}{(k_1 t+c_6)^{{c_2}/{(c_2+2c_4)}}}
+\int_{0} ^t \frac{f(t^{\prime})dt^{\prime}}{
(k_1t^{\prime}+c_6)^{{2(c_2+c_4)}/{(c_2+2c_4)}}},
\]
\[
 \tau_2 = y(k_1t+c_6)^{(c_2-2c_4)/(c_2+2c_4)}
 - \int_{0} ^t (4Bc_3t^{\prime}-c_5)
(k_1t'+c_6)^{{-4c_4}/{(c_2+2c_4)}}dt^{\prime},
\]
\[
\ds F = (c_2 u-c_3)(k_1t+c_6)^{2c_2/(c_2+2c_4)},
\]
\[
\ds G = v(k_1t+c_6)^{2c_4/(c_2+2c_4)}+\int_{0} ^t
\frac{\dot{f}(t^{\prime})dt^{\prime}}{2B(k_1t^{\prime}+
c_6)^ {c_2/(c_2+2c_4)}},
\]
where $k_1=(c_2/2)+c_4 $.

The reduced PDE takes the form
\[
\ba{l}
\ds BF_{\tau_1\tau_1\tau_2}+4BFG_{\tau_1}+2BGF_{\tau_1}
+\frac{(c_2-2c_4)}{2}\tau_2F_{\tau_2}-\frac{c_2}{2}\tau_1F_{\tau_1}-c_2F = 0,
\\[4mm]
\ds F_{\tau_2} = c_2 G_{\tau_1}.
\ea
\]
{\bf Case 2:} $c_1,c_2 = 0$. The similarity variables are
\[
\ba{l}
\ds \tau_1 = x+\int_{0} ^t
\frac{f(t^{\prime})dt^{\prime}}{(c_4t'+c_6)},\\[4mm]
\ds \tau_2 = \frac{y}{(c_4t+c_6)} - \frac{4Bc_3}{c_4 ^2}\log (c_4 t+c_6) -
\frac{(4Bc_3c_6+c_4c_5)}{c_4 ^2(c_4t+c_6)},\\[4mm]
\ds F = u-\frac{c_3}{c_4}\log(c_4t+c_6),\qquad G =
v(c_4t+c_6)+\frac{{f}(t)}{2B}.
\ea
\]
In this case the reduced PDE turns out to be
\be \label{lakshmanan:2.15}
\ba{l}
\ds BF_{\tau_1\tau_1\tau_2}+4BFG_{\tau_1}+2BGF_{\tau_1} -\tau_2
c_4F_{\tau_2}\\[3mm]
\ds \qquad \qquad -\left(4Bc_3+\frac{4Bc_3c_6}{c_4}+c_5\right)
F_{\tau_2}+c_3 = 0,
\qquad  F_{\tau_2} =  G_{\tau_1}.
\ea
\ee
{\bf Case 3:} $c_1,c_4 = 0$. The similarity variables are
\[
\ba{l}
\ds \tau_1 = \frac{x}{(c_2/2)t+c_6}+\int_{0} ^t
\frac{f(t^{\prime})dt^{\prime}} {((c_2/2)t^{\prime}+c_6)^2},\quad
\tau_2 = y((c_2/2)t+c_6)-2Bc_3t^2+c_5t, \\[3mm]
\ds F = (c_2u-c_3)((c_2/2)t+c_6)^2,\quad G = v+\int_{0} ^t
\frac{\dot{f}(t^{\prime})dt'}{2B((c_2/2)t'+c_6)}.
\ea
\]
The reduced PDE takes the form
\[
\ba{l}
\ds
BF_{\tau_1\tau_1\tau_2}+\frac{4BFG_{\tau_1}}{c_2}+\frac{2BGF_{\tau_1}}{c_2}
+\left(c_2c_5c_6+4Bc_3c_6^2\right)G_{\tau_1}\\[3mm]
\ds \qquad \qquad -\frac{\tau_1}{2}F_{\tau_1}
+\frac{\tau_2}{2}F_{\tau_2}-2F = 0,\qquad
F_{\tau_2} = c_2 G_{\tau_1}.
\ea
\]
{Case 4:} $c_1,c_2,c_3 = 0$.  The similarity variables are
\[
\ba{l}
\ds \tau_1 = x+\int_{0}^t
\frac{f(t^{\prime})dt^{\prime}}{c_4t'+c_6},\qquad  \tau_2
= \frac{c_4y}{c_4t+c_6}-\frac{c_5}{c_4t+c_6},
\\[3mm]
\ds F = u,\qquad  G = v(c_4t+c_6)+ \frac{f(t)}{2B}.
\ea
\]
The reduced PDE takes the form
\[
\ba{l}
\ds
BF_{\tau_1\tau_1\tau_2}+\frac{4B}{c_4}FG_{\tau_1}+\frac{2B}{c_4}GF_{\tau_1}
-\tau_2F_{\tau_2}= 0,\\[3mm]
\ds G_{\tau_1} = c_4F_{\tau_2}.
\ea
\]
{\bf Case 5:} $c_1,c_2,c_4 = 0$. The similarity variables are
\[
\ba{l}
\ds \tau_1 = x+\frac{1}{c_6}\int_{0} ^t f(t^{\prime})dt^{\prime},\qquad
\tau_2 = y-\frac{2Bc_3}{c_6}t^2+\frac{c_5}{c_6}t,
\\[3mm]
\ds F = u-\frac{c_3}{c_6} t,\qquad G = v + \frac{f(t)}{2Bc_6}.
\ea
\]
The reduced PDE takes the form
\[
\ba{l}
\ds BF_{\tau_1\tau_1\tau_2}+4BFG_{\tau_1}+2BGF_{\tau_1} +
\frac{c_5}{c_6}F_{\tau_2} +\frac{c_3}{c_6} = 0,\\[3mm]
\ds F_{\tau_2} = G_{\tau_1}.
\ea
\]
{\bf Case 6:} $c_1,c_2,c_6 = 0$. The similarity variables are
\[
\ba{l}
\ds \tau_1 = x+\frac{1}{c_4}\int_{0}
^t\frac{f(t^{\prime})dt^{\prime}}{t^{\prime}},\qquad
\tau_2 = \frac{y}{t}-\frac{4Bc_3}{c_4}\log t-\frac{c_5}{c_4t},
\\[3mm]
\ds F = u-\frac{c_3}{c_4}\log t,\qquad G = vt +
\frac{f(t)}{2Bc_4}.
\ea
\]
The reduced PDE takes the form
\[
\ba{l}
\ds BF_{\tau_1\tau_1\tau_2}+4BFG_{\tau_1}+2BGF_{\tau_1}  -
\frac{4Bc_3}{c_4}F_{\tau_2}-\tau_2 F_{\tau_2}-\frac{c_3}{c_4} = 0,\\[3mm]
\ds F_{\tau_2} = G_{\tau_1}.
\ea
\]
{\bf Case 7:} $c_1,c_3,c_4 = 0$.
The similarity variables are
\[
\ba{l}
\ds \tau_1 = \frac{x}{(c_2/2)t+c_6}+\int_{0} ^t
\frac{f(t^{\prime})dt^{\prime}}{((c_2/2)t'+c_6)^2}, \quad \tau_2 =
y((c_2/2)t+c_5)+c_5t,
\\[3mm]
\ds F = u((c_2/2)t+c_6)^2, \qquad G = v + \int_{0} ^t
\frac{\dot{f}(t^{\prime})dt^{\prime}}{2B((c_2/2)t'+c_6)}.
\ea
\]
The reduced PDE takes the form
\[
\ba{l}
\ds BF_{\tau_1\tau_1\tau_2}+4BFG_{\tau_1}+2BGF_{\tau_1} +
\left[\frac{c_2}{2}\tau_2  +c_5 ^2\right]
F_{\tau_2}-\frac{\tau_1c_2}{2}F_{\tau_1}-c_2F  = 0,
\\[3mm]
\ds F_{\tau_2} = G_{\tau_1}.
\ea
\]

\subsection{Lie symmetries and similarity reduction of eqs.(\ref{lakshmanan:2.11})}

Now the reduced PDE (\ref{lakshmanan:2.11}) in two independent variables can itself be
further analyzed for its symmetry properties by looking at its own
invariance property under the classical Lie algorithm again. The invariance
of the eq.(\ref{lakshmanan:2.11}) leads to the following inf\/initesimal symmetries
\[
\xi_1=c_7\tau_1+c_8,\quad \xi_2=-2c_7\tau_2, \quad \eta_1=-2c_7F, \quad
\eta_2=c_7G+\frac{3c_2c_8}{8Bc_1},
\]
where $c_7$ and $c_8$ are arbitrary constants.  The associated Lie
vector f\/ields are
\[
V_1 = \tau_1\frac{\partial}{\partial\tau_1} -2\tau_2\frac{\partial}{
\partial\tau_2} -2F\frac{\partial}{\partial F}+G\frac{\partial}{\partial G},
\quad V_2 = \frac{\partial}{\partial\tau_1}+
\frac{3c_2}{8Bc_1}\frac{\partial}{\partial G},
\]
leading to the Lie algebra
\[
[V_1,V_2] = V_1.
\]
Solving the associated characteristic equation
\[
\frac{d\tau_1}{c_7 \tau_1 +c_8} = \frac{d\tau_2}{-2c_7 \tau_2} =
\frac{dF}{-2c_7 F } = \frac{dG}{c_7 G+(3c_2c_8/8Bc_1)},
\]
we obtain the similarity variables
\be\label{lakshmanan:2.30}
z=(c_7\tau_1+c_8)\tau_2^{(1/2)}, \quad w_1 = F(c_7\tau_1+c_8)^2,
 \quad w_2 = \frac{c_7G+(3c_2c_8/8Bc_1)}{(c_7\tau_1+c_8)}.
\ee
The associated similarity reduced ODE follows from eqs.(\ref{lakshmanan:2.30})
and (\ref{lakshmanan:2.11}) as
\be\label{lakshmanan:2.31}
\ba{l}
\ds w_1^{\prime\prime\prime}-\frac{2}{z}w_1^{\prime\prime}+
\frac{2}{z^2}w_1^{\prime}
+\frac{8}{c_1^2}w_1w_2^{\prime}+\frac{4}{c_1^2}w_1^{\prime}w_2
-\frac{3c_2+6c_4}{4Bc_1^3}w_1^{\prime}+\frac{3c_6}{2B^2c_1^3}z = 0,
\\[3mm]
\ds w_1^{\prime} = 2z(w_2+zw_2^{\prime}).
\ea
\ee
While the exact solution for eq.(\ref{lakshmanan:2.31}) has not been found for the general
two parameter case, particular solutions can be obtained for
the special one parameter choice.  For example, by choosing
$c_7 = 0$ and redoing the calculations one gets the following solution
\be\label{lakshmanan:2.32}
F = \frac{3c_2}{8Bc_1}\tau_2+I_1, \qquad
G = \frac{3c_2}{8Bc_1}\tau_1+w_2(\tau_2),
\ee
where $I_1$ is an integration constant and $w_2(\tau_2)$ is an arbitrary
function of $\tau_2 $.  Now rewriting eq.(\ref{lakshmanan:2.32}) interms of old variables
one can get a solution for the PDE (\ref{lakshmanan:2.1}). However, the other possibility
$c_8 = 0$ leads to the same similarity reduction (\ref{lakshmanan:2.31}).

Similarly one can analyse each one of the other equations given in
Sec.~\ref{sec:2.3}.
For example let us consider Case~2, eq.(\ref{lakshmanan:2.15}). Now applying the
invariance condition to eq.(\ref{lakshmanan:2.15}) one gets the following
inf\/initesimals
\[
\xi_1=c_7,\qquad \xi_2=\frac{4Bc_8}{c_4}, \qquad \eta_1=c_8,
\qquad \eta_2=0.
\]

Solving the characteristic equation we get the following similarity variables:
\[
z=\tau_1 - \frac{c_4c_7}{4Bc_8}\tau_2, \qquad
w_1 = F-\frac{c_8}{c_7}\tau_1, \qquad w_2 = G.
\]

Under this similarity transformation the reduced ODE takes the form
\be\label{lakshmanan:2.35}
w_1^{'''}+6w_1w_1^{'}+\frac{4c_8}{c_7}zw_1^{'}
+k_1w_1^{'}+\frac{2c_8}{c_7^2}w_1 + k_2 = 0,
\ee
where
\[
k_1 = \left ( \frac{8Bc_8I_1}{c_4c_7}-4c_3-\frac{4c_3c_6}{c_4}- \frac{c_5}{B}
\right ), \qquad
k_2 = - \left ( \frac{4c_3c_8}{c_7}+\frac{8Bc_8^2I_1}{c_4c_7^2} \right ),
\]
and
\[
w_2 = -\frac{c_4c_7}{4Bc_8}w_1+I_1.
\]
Solving eq.(\ref{lakshmanan:2.35}) one gets the solution for the PDE (\ref{lakshmanan:2.15}).

Similarly by choosing $c_7 = 0$, we get the solution
\[
\ba{l}
\ds F = \frac{c_4}{4B}\tau_2+\frac{1}{2}
\left(\frac{I_2}{\tau_1+(4BI_1/c_4)}\right)^2
-\left(\frac{c_3}{c_4}-c_3-\frac{c_3c_6}{c_4}-\frac{c_5}{4B}\right),\\[3mm]
\ds G= \frac{c_4}{4B}\tau_1+I_1,
\ea
\]
where $I_1, I_2$ are integration constants.  Substituting the expressions
for $\tau_2$ and $\tau_1$ from eq.(\ref{lakshmanan:2.15}) one gets a particular solution for
eq.(\ref{lakshmanan:2.1}).

Similarly one can also bring out other particular solutions
for all the other sub-cases in the same manner.

\setcounter{equation}{0}

\section{Lie symmetries and Inf\/inite Dimensional Lie Algebras of the
Zakharov-Strachan equation}

In this section we investigate the symmetries and similarity reductions
associated with another important (2+1)-dimensional generalization of the
NLS equation, introduced originally by Zakharov and studied recently by
Strachan of the form~[22]. Its form reads
\be\label{lakshmanan:3.1}
\ba{l}
\ds 2kq_t = q_{xy}-2q\int \partial _y [p.q]dx, \\[3mm]
\ds -2kp_t = p_{xy}-2p\int \partial _y [p.q]dx.
\ea
\ee
By introducing a potential $v(p,q)$ def\/ined by
\[
v_x(p,q) = 2\partial _y [p.q],
\]
and imposing the algebraic constraints on the f\/ields $p$ and $q$ such
that $q =p^* = \psi $ and choosing $k=i/2 $ eq.(\ref{lakshmanan:3.1}) reduces to
\be\label{lakshmanan:3.3}
i\psi _t = \psi_{xy}+v\psi, \qquad
v_x = 2\partial _y |\psi|^2.
\ee
When $\partial _x = \partial _y $ eq.(\ref{lakshmanan:3.3}) reduces to the NLS equation and
when $\partial _t = 0$, it reduces to a complicated sine-Gordon equation.

It has been shown that in ref.~[23] that eq.(\ref{lakshmanan:3.3}) admits P-property.
Further, the authors have also constructed a new class of localized solutions
called ``induced localized structures"~[24].

To study the invariance properties of the the eq.(\ref{lakshmanan:3.3}) we introduce the
transformation $\psi = a+ib$ so that eq.(\ref{lakshmanan:3.3}) becomes
\be\label{lakshmanan:3.4}
a_t-b_{xy}-bv = 0,\qquad b_t+a_{xy}+av = 0,\qquad
v_x-4aa_y-4bb_y = 0.
\ee
We will investigate the
Lie symmetries of eq.(\ref{lakshmanan:3.4}) under the one parameter $(\varepsilon)$ group of
transformations
\[
\ba{l}
x\longrightarrow X = x+\varepsilon\xi_1(t,x,y,a,b,v),\\[1mm]
y\longrightarrow Y = y+\varepsilon\xi_2(t,x,y,a,b,v),\\[1mm]
t\longrightarrow T = t+\varepsilon\xi_3(t,x,y,a,b,v),\\[1mm]
a\longrightarrow A = a+\varepsilon\phi_1(t,x,y,a,b,v),\\[1mm]
b\longrightarrow B = b+\varepsilon\phi_2(t,x,y,a,b,v),\\[1mm]
v\longrightarrow V = v+\varepsilon\phi_3(t,x,y,a,b,v).
\ea
\]
The inf\/initesimal transformations can be worked out to be
\be\label{lakshmanan:3.6}
\ba{l}
\ds \xi_1 = -\left(\frac{c_1}{2}xt+c_2x-f(t)\right),\\[3mm]
\ds \xi_2 = -\frac{c_1}{2}yt+(c_2-c_3)y-c_5t+c_6,\\[3mm]
\ds \xi_3 = -\left(\frac{c_1}{2}t^2+c_3t+c_4\right),\\[3mm]
\ds \phi_1 = -\frac{c_1}{2}bxy+\frac{c_1}{2}at+c_2a-c_5bx+
\dot {f}(t)by-g(t)b,
\\[3mm]
\ds \phi_2 = \frac{c_1}{2}axy+\frac{c_1}{2}bt+c_2b+
c_5ax -\dot {f}(t)ay+g(t)a,\\[3mm]
\ds \phi_3 = (c_1t+c_3)v+\ddot {f}(t)y-\dot{g}(t),
\ea
\ee
where $c_1,c_2,c_3,c_4,c_5,c_6$ are arbitrary constants and $f(t) $
and $g(t)$ are arbitrary functions of $t$ and dot denotes
dif\/ferentiation with respect to $t$.

\subsection{Lie algebra}

The Lie vector f\/ields associated with the inf\/initesimal transformations can
be written as
\[
V = V_1(f)+V_2(g)+V_3+V_4+V_5+V_6+V_7+V_8,
\]
where $f$ and $g$ are arbitrary functions of $t$ and
\[
\ba{l}
\ds V_1(f) = f(t)\frac{\partial}{\partial x} +\dot{f}(t)by\frac{\partial}{
\partial a}- \dot{f}(t)ay\frac{\partial}{\partial b}+ \ddot{f}(t)y\frac{
\partial}{\partial v},\\[3mm]
\ds V_2(g) = -g(t)b\frac{\partial}{\partial a}+
g(t)a\frac{\partial}{\partial b}
- \dot {g}(t)\frac{\partial}{\partial v},
\\[3mm]
\ds V_3 = -\frac{1}{2}xt\frac{\partial}{\partial x}
-\frac{1}{2}yt\frac{\partial}{\partial y}-
\frac{t^2}{2}\frac{\partial}{\partial t} + \frac{1}{2}(at-bxy)
\frac{\partial}{\partial a} +\frac{1}{2}(bt+axy)\frac{\partial}{\partial b}
+vt\frac{\partial}{\partial v},\\[3mm]
\ds V_4 = -x\frac{\partial}{\partial x} +y\frac{\partial}{\partial y}
+a\frac{ \partial}{\partial a} +b\frac{\partial}{\partial b},\qquad
V_5 = -y\frac{\partial}{\partial y} -t\frac{\partial}{\partial t} +v\frac{
\partial}{\partial v},\\[3mm]
\ds V_6 = -t\frac{\partial}{\partial y} -bx\frac{\partial}{\partial
a} +ax\frac{ \partial}{\partial b},\qquad
V_7 = \frac{\partial}{\partial y}, \qquad
V_8 = \frac{\partial}{\partial t}.
\ea
\]
The corresponding Lie algebra between the vector f\/ields becomes
\[
\ba{l}
\ds \left[V_1(f_1), V_2(f_2)\right] = 0,\quad
 \left[V_2(g_1), V_2(g_2)\right] =
0,\quad \left[V_1(f), V_2(g)\right] = 0,\\[2mm]
\ds \left[V_1,V_3\right] = V_1(-f+3t\dot{f}),\quad
 \left[V_1,V_4\right] = V_2(f-t\dot{f}),
\quad \left[V_1,V_5\right] = \frac{1}{4B}V_2(\dot{f}),
\\[2mm]
\ds \left[V_1,V_6\right] = V_1(\dot{f}), \quad
 \left[V_2,V_3\right] = V_2(g+3t\dot{g}), \quad \left[V_2,V_4\right] = 0,
\quad \left[V_2,V_5\right] = 0, \\[2mm]
\ds  \left[V_2,V_6\right] = V_2(\dot{g}), \quad
\left[V_3,V_4\right] = -2V_4,
\quad
\left[V_3,V_5\right] = V_5, \quad \left[V_3,V_6\right] = 3V_6, \\[2mm]
\ds \left[V_4,V_5\right] = -\frac{3A}{2Bg(t)}V_2,
\quad
\left[V_4,V_6\right] = 4BV_6, \quad \left[V_5,V_6\right] = 0.
\ea
\]

It is interesting to note that the the above algebra does not contain a
{\it Virasoro algebra}, which is typical of integrable
(2+1)-dimensional systems such as the NLS equation of Fokas type and other
integrable systems quoted in the Appendix.

\subsection{Similarity variables and reductions}

The similarity variables can be found by integrating the following
characteristic equation
\[
\ba{l}
\ds \frac{dx}{-(\frac{c_1}{2}xt+c_2x-f(t))} = \frac{dy}{-\frac{c_1}{2}
yt+(c_2-c_3)y-c_5t+c_6} \\[6mm]
\ds =\frac{dt}{-(\frac{c_1}{2}t^2+c_3t+c_4)}
=\frac{da}{-\frac{c_1}{2}bxy+\frac{c_1}{2}at+ c_2a-c_5bx+\dot {f}
(t)by-g(t)b}\\[6mm]
\ds =\frac{db}{\frac{c_1}{2}axy+\frac{c_1}{2}bt+ c_2b+c_5ax-\dot {f}(t)ay+g(t)a} =%
\frac{dv}{c_1vt+c_3v+\ddot {f}(t)y-\dot {g}(t)}.
\ea
\]
Solving the characteristic equation, we obtain the following similarity
transformations:
\[
\tau_1 = \frac{x(t+k_1+k_2)^{n-(1/2)}}{(t+k_1-k_2)^{n+(1/2)}}
+\int_{0} ^{t}\frac{2}{c_1}
\frac{f(t^{\prime})(t^{\prime}+k_1+k_2)^{n-(3/2)}} {(t^{\prime}+k_1-k_2)^{n+(3/2)}}%
dt^{\prime},
\]
\[
\tau_2 = \frac{y(t+k_1+k_2)^{-(n+(1/2))}}{(t+k_1-k_2)^{-n+(1/2)}}
-\int_{0} ^{t}\frac{2}{c_1}
\frac{(c_5t^{\prime}-c_6)(t^{\prime}+k_1+k_2)^{-(n+(3/2))}dt'}
{(t^{\prime}+k_1-k_2)^{-n+(3/2)}},
\]
\[
a = \frac{F_2 \sin U (t+k_1+k_2)^{n-(1/2)}}{(t+k_1-k_2)^{n+(1/2)}},
\qquad
b = \frac{F_2 \cos U (t+k_1+k_2)^{n-(1/2)}}{(t+k_1-k_2)^{n+(1/2)}},
\]
\be \label{lakshmanan:3.11}
F_3 = \frac{c_1}{2}(t+k_1+k_2)(t+k_1-k_2)v
\ee
\[
\qquad +
\frac{\dot{f}(t)(t+k_1-k_2)^{-n+(1/2)}}{(t+k_1+k_2)^{-n-(1/2)}}
\int_{0} ^t
\frac{2(c_5t^{\prime}-c_6)(t^{\prime}+k_1+k_2)^{-n-(3/2)}dt^{\prime}}{
c_1(t^{\prime}+k_1-k_2)^{-n+(3/2)}}
\]
\[
\qquad\!\!
-\frac{f(t)(t+k_1-2nk_2)(t+k_1-k_2)^{-n-(1/2)}}{(t+k_1+k_2)^{-n+(1/2)}}
\!\!\int_{0} ^t\! \frac{2(c_5t^{\prime}-c_6)
(t^{\prime}+k_1+k_2)^{-n-(3/2)}dt^{\prime}}{
c_1(t^{\prime}+k_1-k_2)^{-n+(3/2)}}\!
\]
\[
\qquad + \int_{0} ^t \frac{f(t)(t^{
\prime}+k_1-k_2)^{-n-(1/2)}dt^{\prime}}
{(t^{\prime}+k_1+k_2)^{-n+(1/2)}} \int_{0} ^{t'}
\frac{2(c_5t^{\prime\prime }-c_6)(t^{\prime\prime }+k_1+k_2)^{-n-(3/2)}dt^
{\prime\prime}}{
c_1(t^{\prime\prime }+k_1-k_2)^{-n+(3/2)}}
\]
\[
\qquad - \int_{0} ^t \frac{
f(t')(t'+k_1+2nk_2)(t'+k_1-2nk_2)(t^{\prime}+k_1-k_2)^{-n-(3/2)}dt^{\prime}}
{(t^{\prime}+k_1+k_2)^{-n+(3/2)}}
\]
\[
\qquad \times
\int_{0} ^{t^\prime}\frac{2(c_5t^{\prime \prime}
-c_6)(t^{
\prime \prime}+k_1+k_2)^{-n-(3/2)}dt^{\prime \prime }}
{c_1(t^{\prime \prime}+k_1-k_2)^{-n+(3/2)}}
\]
\[
\qquad  +
 \int_{0} ^t \frac{2f(t')(t'+k_1-2nk_2)(c_5t'-c_6)dt'}{
c_1(t'+k_1+k_2)^2(t'+k_1-k_2)^2}
 -\frac{2f(t)(c_5t-c_6)}{c_1 (t+k_1+k_2)(t+k_1-k_2)}
\]
\[
\qquad +
 \int_{0} ^t \frac{2c_5f(t')dt'}{c_1(t'+k_1+k_2)(t'+k_1-k_2)}
 - \int_{0} ^t \frac{4f(t')(t'+k_1)(c_5t'-c_6)dt'}{%
c_1(t'+k_1+k_2)^2(t'+k_1-k_2)^2}  - g(t) ,
\]
where
\[
k_1 = \frac{c_3}{c_1},\qquad k_2 = \frac{\sqrt {c_3 ^2 -2c_1
c_4}}{c_1}, \qquad n=\frac{2c_2-c_3}{2\sqrt {c_3 ^2 -2c_1 c_4}}
\]
and
\[
U = - \int_{0} ^t \left[\int_{0} ^{t^{\prime}}
\frac{2f(t^{\prime\prime})(t^{\prime
\prime}+k_1+k_2)^{n-(3/2)}} {c_1(t^{\prime\prime}+k_1-k_2)^{n+(3/2)}}
dt^{\prime\prime}\right]
\]
\[
\qquad \times\left[\int_{0} ^{t^{\prime}}\frac{2(c_5
t^{\prime\prime}-c_6)(t^{\prime\prime}+k_1+k_2)^{-(n+(3/2))}} {
c_1(t^{\prime\prime}+k_1-k_2)^{-n+(3/2)}}dt^{\prime\prime}\right]dt^{\prime}
\]
\[
\qquad -\tau_2 \int_{0} ^t
\left[ \int_{0} ^{t^{\prime}}\frac{2f(t^{\prime
\prime})(t^{\prime\prime}+k_1+k_2)^{n-(3/2)}} {c_1(t''+k_1-k_2)^{n+(3/2)}}
dt^{\prime\prime}\right]dt^{\prime}+\tau_1 \tau_2 t
\]
\[
\qquad +\tau_1 \int_{0} ^{t}\left
[\int_{0} ^{t^{\prime}}\frac{2(c_5
t^{\prime\prime}-c_6)(t^{\prime\prime}+k_1+k_2)^{-(n+(3/2))}} {
c_1(t^{\prime\prime}+k_1-k_2)^{-n+(3/2)}}dt^{\prime\prime}\right]dt^{\prime}
\]
\[
\qquad - \int_{0} ^t \frac{2c_5 (t^{\prime}+k_1-k_2)^{n-(1/2)}}{c_1
(t^{\prime}+k_1+k_2)^{n+(1/2)}}
\left[ \int_{0} ^{t^{\prime}}\frac{
2f(t^{\prime\prime})(t^{\prime\prime}+k_1+k_2)^{n-(3/2)}
dt^{\prime\prime}}{c_1(t^{\prime
\prime}+k_1-k_2)^{n+(3/2)}} \right]dt^{\prime}
\]
\[
\qquad+\tau _1 \int_{0} ^{t}
\frac{2c_5(t^{\prime}+k_1-k_2)^{n-(1/2)}} {c_1
(t^{\prime}+k_1+k_2)^{n+(1/2)}}dt^{\prime}+
\int_{0} ^{t}\frac{4f(t')(c_5t^{\prime}-c_6)}{
c_1 ^2 (t^{\prime}+k_1+k_2)^2(t^{\prime}+k_1-k_2)^2}dt^{\prime}
\]
\[
\qquad -\frac {f(t)(t+k_1-k_2)^{-(n+(1/2))}}{(t+k_1+k_2)^{-n+(1/2)}}
 \int_{0} ^{t}\frac{
4(c_5 t^{\prime}-c_6)(t^{\prime}+k_1+k_2)^{-(n+(3/2))}} {c_1 ^2
(t^{\prime}+k_1-k_2)^{-n+(3/2)}}dt^{\prime}
\]
\[
\qquad -\int_{0} ^{t}\frac{2f(t^{\prime})(t^{\prime}+k_1+2nq)(t^{
\prime}+k_1-k_2)^{-(n+(3/2))}}{c_1(t^{\prime}+k_1+k_2)^{-n+(1/2)}}
\]
\[
\qquad \times\left[\int_{0}
^{t^{\prime}}\frac{2(c_5
t^{\prime\prime}-c_6)(t^{\prime\prime}+k_1+k_2)^{-(n+(3/2))}} {
c_1(t^{\prime\prime}+k_1-k_2)^{-n+(3/2)}} dt^{\prime\prime}\right]dt'
\]
\[
\qquad -\frac{2\tau_2 f(t)(t+k_1-k_2)^{-(n+(1/2))}}{c_1
(t+k_1+k_2)^{-n+(1/2)}}
\]
\[
\qquad -
\tau_2 \int_{0} ^{t}\frac{2f(t^{\prime})(t^{\prime}+k_1+2nq)(t^{
\prime}+k_1-k_2)^{-(n+(3/2))}}{c_1
(t^{\prime}+k_1+k_2)^{-n+(3/2)}}dt^{\prime}
\]
\[
\qquad + \int_{0} ^{t}\frac{2g(t^{\prime})
dt^{\prime}}{c_1(t^{\prime}+k_1+k_2)(t^{\prime}+k_1-k_2)}
+F_1,
\]
where $F_1$, $F_2$ and $F_3$ are arbitrary functions of $\tau_1$ and
$\tau_2$.  Under this set of similarity transformations
eq.(\ref{lakshmanan:3.4}) takes the form
\be \label{lakshmanan:3.12}
\hspace*{-13.5pt}
\ba{l}
\ds F_{2\tau_1\tau_2}+\frac{2c_2}{c_1}\tau_1 F_2F_{1\tau_1} +\frac{2(c_3-c_2)}{c_1}%
\tau_2F_2F_{1\tau_2}
-F_2F_{1\tau_1} F_{1\tau_2}
-\frac{2c_4}{c_1}\tau_1 \tau_2 F_{2}+2F_2F_3 = 0,\\[3mm]
\ds F_2 F_{1\tau_1 \tau_2}+F_{2\tau_1}F_{1\tau_2}+F_{1\tau_1}F_{2\tau_2}
- \frac{2c_2}{c_1}\tau_1
F_{2\tau_1}  -\frac{2(c_3-c_2)}{c_1}\tau_2 F_{2\tau_2} -
\frac{2c_2}{c_1}F_2 = 0, \\[3mm]
\ds F_{3\tau_1}-2c_1F_2F_{2\tau_2} = 0.
\ea\!\!\!\!\!
\ee

\subsection{Subcases}
\label{sec:3.3}
Besides the above general similarity reductions, one can f\/ind a number of
special reductions corresponding to lesser parameter symmetries by choosing
some of the arbitrary parameter $c_1$, $c_2$, $c_3$, $c_4$ and $c_5$
and arbitrary functions $f(t)$ and $g(t)$ to be zero. Important nontrivial
cases are given below.

\noindent
{\bf Case 1:} $c_1 = 0$.
\[
\tau_1 = \frac{x}{(c_3t+c_4)^{c_2/c_3}}+\int_{0} ^t \frac{f(t')dt'}
{(c_3t'+c_4)^{(c_2+c_3)/c_3}},
\]
\[
\tau_2 = y(c_3t+c_4)^{(c_2-c_3)/c_3}-\int_{0} ^t
(c_5t'-c_6)(c_3t'+c_4)^{(c_2-2c_3)/c_3}dt',
\]
\[
a = F_2 \sin U (c_3t+c_4)^{-c_2/c_3},\qquad
b = F_2 \cos U (c_3t+c_4)^{-c_2/c_3},
\]
\[
U = -c_5 \int_{0} ^t (c_3t'+c_4)^{{(c_2-c_3)}/{c_3}}\left
[ \int_{0} ^{t'} \frac{f(t'')dt''}
{(c_3t''+c_4)^{(c_2+c_3)/c_3}}\right ]dt'
\]
\[
\qquad +c_5 \tau_1 \int_{0} ^t
(c_3t'+c_4)^{(c_2-c_3)/c_3}dt'
\]
\[
\qquad- \int_{0} ^t \dot {f}(t')(c_3t'+c_4)^{-{c_2}/{c_3}} \left [
\int_{0} ^{t'} (c_5t''-c_6)(c_3t''+c_4)^{(c_2-2c_3)/c_3}dt''
\right ] dt'
\]
\[
\qquad -\tau_2 \int_{0} ^t \dot {f}(t')(c_3t'+c_4)^{-c_2/c_3}dt'
+\int_{0} ^t \frac{g(t')dt'}{(c_3t'+c_4)}+F_1.
\]
Under this similarity transformation the reduced PDE takes the form
\[
\ba{l}
\ds F_{2\tau_1\tau_2}+c_2\tau_1 F_2F_{1\tau_1} +(c_3-c_2)%
\tau_2F_2F_{1\tau_2}-F_2F_{1\tau_1} F_{1\tau_2}
-2F_2F_3 = 0,\\[1mm]
\ds F_2 F_{1\tau_1 \tau_2}+F_{2\tau_1}F_{1\tau_2}+F_{1\tau_1}F_{2\tau_2}
-c_2F_{2} -c_2\tau_1 F_{2\tau_1} + (c_2-c_3)\tau_2F_{2\tau_2} = 0,
\\[1mm]
\ds F_{3\tau_1}-F_2F_{2\tau_2} = 0.
\ea
\]
{\bf Case 2:} $c_1,c_2,c_3 = 0$.
\[
\tau_1 = x+\frac{1}{c_4}\int_{0} ^t{f(t^{\prime})dt^{\prime}},
 \quad \tau_2 = y-\frac{c_5t^2}{2c_4}+\frac{c_6t}{c_4}, \quad
a = F_2 \sin U, \quad b = F_2 \cos U,
\]
\[
v = -\frac{c_5}{2c_4}t^2 \dot {f}(t) + \frac{c_5}{c_4}tf(t) -\frac{c_5}{c_4}
\int_{0} ^t f(t')dt' +\frac{c_6}{c_4}t\dot{f}(t)
-\frac{c_6}{c_4}f(t)-\tau_2 \dot{f}(t) +\frac{g(t)}{c_4}+F_3.
\]
and
\[
U = -\frac{c_5}{c_4^2}\int_{0} ^t \left [\int_{0} ^{t'}
f(t'')dt''\right ]dt' +\frac{c_5}{c_4}\tau_1 t
-\frac{c_5}{2c_4^2}\left [t^2 f(t)-2 \int_{0} ^t t'f(t')dt'\right ]
\]
\[
\qquad +\frac{c_6}{c_4^2}\left[tf(t)-\int_{0} ^t
f(t')dt'\right] -\frac{\tau_2 f(t)}{c_4}+\frac{1}{c_4}\int_{0} ^t
g(t')dt' +F_1.
\]
The reduced PDE takes the form
\[
\ba{l}
\ds F_2F_{1\tau_1\tau_2}+F_{2\tau_1}F_{1\tau_2}+F_{2\tau_2}F_{1\tau_1}
+\frac{c_6}{c_4}F_{2\tau_2} = 0,\\[3mm]
\ds F_{2\tau_1\tau_2}-F_2F_{1\tau_1}F_{1\tau_2}-\frac{c_6}{c_4}F_2F_{1\tau_2}
- \frac{c_5}{c_4}\tau_1F_2+F_2F_3  = 0,\\[3mm]
\ds F_{3\tau_1}-4F_2F_{2\tau_2} = 0.
\ea
\]
{\bf Case 3:} $c_1,c_2,c_4 = 0$.
\[
\tau_1 = x+\frac{1}{c_3}\int_{0} ^t \frac{f(t^{\prime})dt^{\prime}}{t'},
 \quad \tau_2 = \frac{y}{t}-\frac{c_5}{c_3}\log t-\frac{c_6}{c_3t}, \quad
a = F_2 \sin U, \quad b = F_2 \cos U,
\]
\[
v = -\frac{\dot{f}(t)}{c_3}\log t ^{{c_5}/{c_3}}+\frac{f(t)}{c_3t}
\log t^{{c_5}/{c_3}} - \frac{c_5}{c_3^2t}\int_{0} ^t \frac{f(t')dt'}{t'}
+\frac{c_5}{c_3^2 t}f(t) - \frac{c_6}{c_3^2 t}\dot{f}(t)
\]
\[
\qquad \qquad -\frac{\tau_2}{c_3}
\dot{f}(t)
+\frac{\tau_2}{c_3t}f(t)+\frac{g(t)}{c_3t}+\frac{F_3}{t},
\]
and
\[
U = -\frac{c_5}{c_3^2}\int_{0} ^t \frac{1}{t'} \left [\int_{0} ^{t'}
\frac{ f(t'')}{t^{''}}dt''\right ]dt'
+\frac{c_5\tau_1}{c_3}\log t -\frac{f(t)}{c_3}\log t^{{c_5}/{c_3}}
+\frac{c_5}{c_3^2}\int_{0} ^t \frac{f(t')dt'}{t'}
\]
\[
\qquad  - \frac{c_6f(t)}{c_3^2 t}
-\frac{c_6}{c_3^2}\int_{0} ^t \frac{f(t')}{t'^2}dt' - \frac{\tau_2
f(t)}{c_3} + \int_{0} ^t \frac{g(t')dt'}{c_3 t'} + F_1.
\]
The reduced PDE takes the form
\[
\ba{l}
\ds F_2F_{1\tau_1\tau_2}+F_{1\tau_1}F_{2\tau_2}-\frac{c_5}{c_3}F_{2\tau_2}
-\tau_2F_{2\tau_2} = 0,\\[3mm]
\ds F_{2\tau_1\tau_2}-F_2F_{1\tau_1}F_{1\tau_2}+\frac{c_5}{c_3}F_2F_{1\tau_2}
+\tau_2F_2F_{1\tau_2}-\frac{c_5}{c_3}\tau_1F_2+F_2F_3 = 0,\\[3mm]
\ds F_{3\tau_1}-4F_2F_{2\tau_2} = 0.
\ea
\]
{\bf Case 4:} $c_1,c_2,c_6 = 0$.
\[
\tau_1 = x+\int_{0} ^t\frac{f(t')}{(c_3t'+c_4)}dt^{\prime}, \;\;\;
\tau_2 = \frac{y}{(c_3t+c_4)}-log(c_3t+c_4)^{{c_5}/{c_3^2}} -
\frac{c_4c_5}{c_3^2(c_3t+c_4)},
\]
\[
a = F_2 \sin U, \qquad b = F_2 \cos U,
\]
\[
v = -\frac{1}{(c_3t+c_4)}\left [ \dot {f}(t)(c_3t+c_4)\log
(c_3t+c_4)^{c_5/c_3^2} -c_3f(t) \log(c_3t+c_4)^{c_5/c_3^2}
\right.
\]
\[
\left.
\qquad +c_5 \int_{0} ^t \frac {f(t')}{(c_3t'+c_4)}dt'
-\frac{c_5}{c_3}f(t) +\frac{c_4c_5}{c_3^2}\dot {f}(t)+\tau_2 \dot {f}(t)
(c_3t+c_4) - c_3\tau_2 f(t) \right ]
\]
\[
U = -c_5 \int_{0} ^t \frac{1}{(c_3t'+c_4)} \left [ \int_{0} ^ {t'}
\frac{ f(t'')}{(c_3t''+c_4)}dt'' \right ] dt' + \frac{\tau_1 c_5}{c_3}
\log(c_3t+c_4)
\]
\[
\qquad -f(t)\log (c_3t+c_4)^{{c_5}/{c_3^2}}+\frac{c_5}{c_3}\int_{0} ^t
\frac{f(t')}{(c_3t'+c_4)}dt'
 - \frac{c_4c_5f(t)}{c_3^2(c_3t+c_4)}
\]
\[
\qquad -\frac{c_4c_5}{c_3}\int_{0} ^t \frac{f(t')}{(c_3t'+c_4)^2}dt'
- \tau_2 f(t).
\]
The reduced PDE takes the form
\[
\ba{l}
\ds F_2F_{1\tau_1\tau_2}+F_{1\tau_2}F_{2\tau_1}+F_2\tau_2F_{1\tau_1}
-c_3\tau_2F_{2\tau_2} -\frac{c_5}{c_3}F_{2\tau_2} = 0,\\[3mm]
\ds F_{2\tau_1\tau_2}-F_2F_{1\tau_1}F_{1\tau_2}+\frac{c_5}{c_3}F_2F_{1\tau_2}
+c_3\tau_2F_2F_{1\tau_2}+c_5\tau_1F_2+F_2F_3 = 0,\\[3mm]
\ds F_{3\tau_1}-4F_2F_{2\tau_2} = 0.
\ea
\]

\subsection{Lie symmetries of eq.(\ref{lakshmanan:3.12})}

Applying the Lie algorithm again to the eq.(\ref{lakshmanan:3.12}), one gets the
inf\/initesimals as
\be \label{lakshmanan:3.21}
\ba{l}
\ds \xi_1 = -c_7\tau_1+\frac{c_1^2c_8}{(2c_2c_3-c_1c_4-2c_2^2)},\qquad
\xi_2 = c_7\tau_2,\\[5mm]
\ds  \phi_1 = \frac{2c_1c_2c_8\tau_2}{(2c_2c_3-c_1c_4-2c_2^2)}+c_9,
\qquad \phi_2 = c_7 w_2,\qquad  \phi_3 = c_8 \tau_2,
\ea
\ee
where $c_7$, $c_8$ and $c_9$ are arbitrary constants.
Solving the characteristic equation associated with inf\/initesimal symmetries,
(\ref{lakshmanan:3.21}), we get the following similarity variables
\[
z = \tau_2\left(c_7\tau_1 - \frac{c_1^2 c_8}{(2c_2c_3-c_1c_4-2c_2^2)}
\right),
\]
\[
w_1 = F_1 - \frac{2c_1c_2c_8}{c_7(2c_2c_3-c_1c_4-2c_2^2)}\tau_2 -
\frac{c_9}{c_7}\log \tau_2,
\]
\[
w_2 = \frac{F_2}{\tau_2}, \qquad w_3 = F_3 - \frac{c_8}{c_7}\tau_2.
\]
Under this similarity transformation one can reduce the PDE (\ref{lakshmanan:3.12}) into an
ODE of the form
\be \label{lakshmanan:3.23}
\ba{l}
\ds w_2''+\frac{2c_3}{c_1c_7}w_2w_1'-\frac{c_9}{c_7z}w_2w_1'-w_2w_1^2+
\frac{2(c_3-c_2)c_9}{c_1c_7z}w_2\\[4mm]
\ds \qquad \qquad \qquad -\frac{2c_4}{c_1c_7}w_2+\frac{2}{z}w_2w_3
+\frac{2}{z}w_3' = 0,\\[4mm]
\ds w_1''w_2+2w_1'w_2'+\frac{2}{z}w_1'w_2-\frac{2c_3}{c_1c_7}w_2'+\frac{c_9}{c_7}w_2'
-\frac{2c_3w_2}{c_1c_7z} = 0,\\[4mm]
\ds  w_3' = \frac{2c_1}{c_7}w_2(w_2+zw_2').
\ea
\ee

\subsection{Subcases}

Eventhough it is very dif\/f\/icult to f\/ind the general solution of the
eq.(\ref{lakshmanan:3.23})
one can get particular solutions from out of the inf\/initesimal symmetries by
choosing some of the arbitrary constants to be zero.  For example, by choosing
$c_7$ as zero $(c_8, c_9 \neq 0)$ one gets the similarity variables as
\[
z = \tau_2, \quad w_1 = F_1-\frac{2c_2}{c_1}\tau_1\tau_2 -
\frac{c_9k}{c_1^2c_8}\tau_1,\quad
 w_2 = F_2, \quad w_3 = F_3+\frac{k}{c_1^2}\tau_1\tau_2,
\]
where $ k = (2c_2c_3-c_1c_4-2c_2^2) $ and $w_1,w_2$ and $w_3$ are arbitrary
functions of $ \tau_1$ and $\tau_2$.  Under this similarity
transformation the PDE (\ref{lakshmanan:3.12})
gets reduced to an ODE of the following form,
\be \label{lakshmanan:3.25}
\ba{l}
\ds w_2'\left[\frac{4c_2-2c_3}{c_1}z+\frac{c_9k}{c_1^2c_8}\right] = 0,
\qquad w_2w_2'+\frac{kz}{c_1^3 } = 0,\\[4mm]
\ds
\frac{2(c_3-c_2)}{c_1}zw_2w_1'-w_2w_1'
\left[\frac{2c_2}{c_1}z+\frac{c_9k}{c_1^2c_8}\right]
+2w_2w_3 = 0.
\ea
\ee
A simple solution can be obtained from (\ref{lakshmanan:3.25}) by restricting $k=0$, as
\[
w_1 = \frac{c_1}{2c_2-c_3}\int_{0} \frac{w_3}{z}dz +I_2,
\qquad
w_2 = I_1,
\]
where $ w_3 $ is arbitrary.

Similarly for the case $ c_7 = 0$ one ends up with the following ODE:
\[
\ba{l}
\ds w_2''+\frac{2c_3}{c_1}w_2w_1'+\frac{c_9}{c_7z}w_1'w_2-w_2w_1'^2
-\frac{2c_4}{c_1}w_2+\frac{2}{z}w_2w_3 = 0,\\[3mm]
\ds w_2w_1''+2w_1'w_2'-\frac{c_9}{c_1z}w_2'-\frac{2c_3}{c_1}w_2' = 0,
\qquad
w_3' -\frac{2c_1}{z}w_2w_2' = 0.
\ea
\]

One can extend the same analysis for all sub-cases mentioned in
Sec.~\ref{sec:3.3} and bring out particular solutions.

\section{Conclusions}

In this paper we have carried out a detailed invariance analysis of two
dif\/ferent nonlinear evolution equations in (2+1)-dimensions, namely, (i)
breaking soliton equation and (ii) (2+1) NLS equation introduced by
Zakharov, which attracted considerable attention in the recent literature
and pointed out that the above two equations do not admit Virasoro type
algebras even though they are integrable. We have also brief\/ly reviewed the
existence of Kac-Moody-Virasoro algebras in other integrable systems.
The fuller implication of the absence of the Kac-Moody-Virasoro type
subalgebras in both the systems and their connection with integrability
deserves much further study. As far as our knowledge goes no one has pointed
out in the literature that any nonintegrable system admits Kac-Moody-Virasoro
type algebras. Thus from our studies we have also concluded that one can not
distinguish the integrable systems with the existence of Kac-Moody-Virasoro
algebras. Currently we are investigating the possible new similarity
reductions through non-classical and direct methods of Clarkson and Kruskal.

\medskip

\noindent {\bf Acknowledgements:} The work forms part of a
Department of Science and Technology, Government of India research project.

\section{Appendix}
In the following we brief\/ly summarize the existence of Virasoro type algebras
in other important integrable (2+1)-dimensional nonlinear systems.

\appendix

\renewcommand{\theequation}{A.\arabic{equation}}
\setcounter{equation}{0}

\section{Nizhnik-Novikov-Veselov (NNV) equation}
A symmetric generalization of the KdV equation in (2+1)-dimensions is the
NNV equation~[26]
\be\label{lakshmanan:A.1}
\ba{l}
u_t+u_{xxx}+u_{yyy}+u_x+u_y  =  3(uv)_x+3(uq)_y, \\[1mm]
 u_x  =  v_y, \qquad  u_y  =  q_x.
\ea
\ee
It has been shown that eq.(\ref{lakshmanan:A.1}) admits weak Lax pair~[27], Painlev\'{e}
property and dromion solutions~[28].  Further,
eq.(\ref{lakshmanan:A.1}) admits the following inf\/inite dimensional Lie vector
f\/ields of the form~[17]
\[
V=V_1(f)+V_2(g)+V_3(h),
\]
where
\[
\ba{l}
\ds V_1(f) = \frac{x}{3}\dot{f}(t)\frac{\partial}{\partial
x}+\frac{y}{3}\dot{f}(t)\frac{\partial}{\partial y}
+f(t)\frac{\partial}{\partial
t}-\frac{2}{3}u\dot{f}(t)\frac{\partial}{\partial u}
\\[4mm]
\ds \qquad
+\left(\frac{2}{9}\dot{f}(t)-\frac{2}{3}v\dot{f}(t)-
\frac{1}{9}x\ddot{f}(t)\right)
\frac{\partial}{\partial v}  +\left(\frac{2}{9}\dot{f}(t)-
\frac{2}{3}q\dot{f}(t)-\frac{1}{9}y\ddot{f}(t)\right)
\frac{\partial}{\partial q},\\[4mm]
\ds V_2(g) = g(t)\frac{\partial}{\partial x}-
\frac{1}{3}\dot{g}(t)\frac{\partial}{\partial v},
\qquad
V_3(g) = h(t)\frac{\partial}{\partial y}-
\frac{1}{3}\dot{h}(t)\frac{\partial}{\partial q}.
\ea
\]
where $ f(t)$, $g(t)$ and $ h(t) $ are arbitrary functions of $ t $ and
dot denotes dif\/ferentiation with respect to $t$.

The associated Lie algebra between these vector f\/ields become
\[
\ba{l}
\ds \left[V_1(f_1), V_1(f_2)\right] =
V_1(f_1\dot{f}_2-f_2\dot{f}_1),\qquad
\left[V_2(g_1), V_2(g_2)\right]  =  0,\\[3mm]
\ds \left[V_3(h_1), V_3(h_2)\right]  =  0,\qquad
\left[V_1(f), V_2(g)\right]  =
V_2\left(f\dot{g}-\frac{1}{3}g\dot{f}\right),\\[3mm]
\ds \left[V_1(f), V_3(h)\right]  =
V_3\left(f\dot{h}-\frac{1}{3}h\dot{f}\right),\qquad
\left[V_2(g), V_3(h)\right]  =  0,
\ea
\]
which is obviously an inf\/inite dimensional Lie algebra of symmetries.  A
Virasoro-Kac-Moody type subalgebra is immediately obtained by restricting
the arbitrary functions $f$, $g$ and $ h $ to Laurent polynomials so that
we have the commutators
\[
\ba{l}
\ds \left[V_1(t^n), V_1(t^m)\right]  =  (m-n)V_1(t^{n+m-1}),\quad
\left[V_1(t^n), V_2(t^m)\right]  =  \left(m-\frac{1}{3}
n\right)V_2(t^{n+m-1}),
\\[3mm]
\ds \left[V_1(t^n), V_3(t^m)\right] = \left(m-\frac{1}{3}
n\right)V_3(t^{n+m-1}),\quad
\left[V_2(t^n),V_2(t^m)\right]  =  0,\\[3mm]
\ds \left[V_3(t^n),V_3(t^m)\right]  =  0,\quad
\left[V_2(t^n),V_3(t^m)\right]  =  0.
\ea\!
\]

\renewcommand{\theequation}{B.\arabic{equation}}
\setcounter{equation}{0}

\section{Generalized nonlinear Schr\"{o}dinger equation introduced
by Fokas}
Recently Fokas has introduced a (2+1)-dimensional generalized nonlinear
Schr\"{o}dinger equation of the form~[29]
\be \label{lakshmanan:B.1}
\ba{l}
iq_t-(\alpha-\beta)q_{xx}+(\alpha+\beta)q_{yy}
-2\lambda q\left[(\alpha+\beta)v-(\alpha-\beta)u\right]  =  0,\\[1mm]
v_x  =  |q|^2_y,\qquad u_y  =  |q|^2_x.
\ea
\ee

Eq.(\ref{lakshmanan:B.1}) is a symmetric generalization of a (1+1)-dimensional
NLS equation. Inte\-res\-ting\-ly it includes the following three
important systems:

(i) $\alpha = \beta = 1/2$: Simplest complex scalar equation in
(2+1)-dimensions;

(ii) $\alpha = 0$, $\beta = 1$: Davey-Stewartson equation I (DSI);

(iii) $\alpha = 1$, $\beta = 0$: Davey-Stewartson equation III
(DSIII).

\noindent
By introducing the transformation $ q = a+ib $ Eq.(\ref{lakshmanan:B.1}) can be
rewritten as
\be \label{lakshmanan:B.2}
\ba{l}
\ds a_t-(\alpha - \beta)b_{xx}+(\alpha +\beta)b_{yy}-2\lambda(\alpha+\beta)bv
+2\lambda(\alpha-\beta)bu = 0,\\[1mm]
\ds b_t+(\alpha - \beta)a_{xx}-(\alpha +\beta)a_{yy}+2\lambda(\alpha+\beta)av
-2\lambda(\alpha-\beta)au = 0,\\[1mm]
v_x-2aa_y-2bb_y = 0, \qquad u_y-2aa_x-2bb_x = 0.
\ea
\ee

Recently Radha and Lakshmanan~[30] have investigated the system
(\ref{lakshmanan:B.2}) and shown that it admits P-property and constructed
multidromion and localized breather solutions. Eq.(\ref{lakshmanan:B.2}) admits Lie
vector f\/ields of the form~[17]
\[
V = V_1(f)+V_2(g)+V_3(h)+V_4(l)+V_5(m),
\]
where
\[
V_1(f)  =  \frac{x}{2}\dot{f}(t)\frac{\partial}{\partial x}
+\frac{y}{2}\dot{f}(t)\frac{\partial}{\partial y}
+f(t)\frac{\partial}{\partial t}
-\left(\frac{a}{2}\dot{f}(t)-\frac{b}{8A}x^2\ddot{f}(t)
+\frac{b}{8B}y^2\ddot{f}(t)\right)\frac{\partial}{\partial a}
\]
\[
\qquad  -\left(\frac{b}{2}\dot{f}(t)+\frac{a}{8A}x^2\ddot{f}(t)
-\frac{a}{8B}y^2\ddot{f}(t)\right)\frac{\partial}{\partial b}
  -\left(v\dot{f}(t)
-\frac{1}{16B^{2}C}y^{2}\ddot{f}(t)\right)\frac{\partial}{\partial v}
\]
\[
\qquad  -\left(u\dot{f}(t)
+\frac{1}{16A^{2}C}x^{2}\ddot{f}(t)\right)\frac{\partial}
{\partial u},
\]
\[
V_2(g)  =  g(t)\frac{\partial}{\partial x}
+\frac{1}{2A}bx\dot{g}(t)\frac{\partial}{\partial a}
-\frac{1}{2A}ax\dot{g}(t)\frac{\partial}{\partial b}
-\frac{1}{4A^{2}C}x\ddot{g}(t)\frac{\partial}{\partial u},
\]
\[
V_3(h)  =  h(t)\frac{\partial}{\partial y}
-\frac{1}{2B}by\dot{h}(t)\frac{\partial}{\partial a}
+\frac{1}{2B}ay\dot{h}(t)\frac{\partial}{\partial b}
-\frac{1}{4B^{2}C}y\ddot{h}(t)\frac{\partial}{\partial v},
\]
\[
V_4(l)  =  -bl(t)\frac{\partial}{\partial a}
+al(t)\frac{\partial}{\partial b}
+\frac{1}{2AC}\dot{l}(t)\frac{\partial}{\partial u},
\qquad
V_5(m)  =  m(t)\frac{\partial}{\partial u}
+\frac{B}{A}m(t)\frac{\partial}{\partial v},
\]
where $f$, $g$, $h$, $l$, $m$ are arbitrary functions of $t$ and
$A=(\alpha-\beta)$, $B=(\alpha+\beta)$ and $c=\lambda.$  The nonzero
commutation relations between the Lie vector f\/ields are
\[
\ba{l}
\ds \left[V_1(f_1), V_1(f_2)\right] =
V_1(f_1\dot{f}_2-f_2\dot{f}_1),\qquad
\left[V_2(g_1), V_2(g_2)\right]  =  -\frac{1}{2A}
V_4(g_1\dot{g}_2-g_2\dot{g}_1),\\[3mm]
\ds \left[V_3(h_1), V_3(h_2)\right]  =  -\frac{1}{2B}
V_4(h_1\dot{h}_2-h_2\dot{h}_1),\qquad
\left[V_1(f), V_2(g)\right]  =  V_2\left(f\dot{g}
-\frac{g\dot{f}}{2}\right),\\[3mm]
\ds \left[V_1(f), V_3(h)\right]  =  V_3\left(f\dot{h}
-\frac{h\dot{g}}{2}\right), \qquad
\left[V_1(f), V_4(h)\right]  =  V_4(f\dot{l}),\\[3mm]
\ds \left[V_1(f), V_5(m)\right]  =  V(m\dot{f}+f\dot{m}).
\ea
\]
By restricting the arbitrary functions $f$, $g$, $h$, $l$ and
$m$ to be polynomials
in $t$ one can get Kac-Moody-Virasoro type subalgebras of the form
\[
\ba{l}
\ds \left[V_1(t^n), V_1(t^m)\right]  =  (m-n)V_1(t^{n+m-1}),\quad
\left[V_2(t^n), V_2(t^m)\right]  =  \frac{-(m-n)}{2A}V_4(t^{n+m-1}),
\\[3mm]
\ds \left[V_3(t^n), V_3(t^m)\right] =
\frac{-(m-n)}{2B}V_4(t^{n+m-1}), \quad
\left[V_1(t^n), V_2(t^m)\right]  = (m-\frac{n}{2})V_2(t^{n+m-1}) ,
\\[3mm]
\ds \left[V_1(t^n), V_3(t^m)\right]  =  (m-\frac{n}{2})V_3(t^{n+m-1}),\quad
\left[V_1(t^n), V_5(t^m)\right]  =  (m+n)V_5(t^{n+m-1}).
\ea
\]

\renewcommand{\theequation}{C.\arabic{equation}}
\setcounter{equation}{0}

\section{(2+1)-dimensional sine-Gordon equation}

The (2+1)-dimensional integrable sine-Gordon equation introduced by
Konopelchenko and Rogers~[31] in appropriate variables has the form
\be \label{lakshmanan:C.1}
\theta_{xyt}+\frac{1}{2}\theta_y\rho_x
+\frac{1}{2}\theta_x\rho_y  =  0,\qquad
\rho_{xy}-\frac{1}{2}(\theta_x\theta_y)_t  =  0.
\ee

Recently Radha and Lakshmanan have studied singularity structure and localized
solutions of the eq.(\ref{lakshmanan:C.1}) and shown that eq.(\ref{lakshmanan:C.1}) admits
P-property~[32].
Eq.(\ref{lakshmanan:C.1}) admits the following Lie vector f\/ields~[17]
\[
V = V_1(f)+V_2(g)+V_3(h)+V_4(l)+V_5(N)
\]
where
\[
\ba{l}
\ds V_1 =  f(x)\frac{\partial}{\partial x},\qquad
V_2  =  g(y)\frac{\partial}{\partial y},\qquad
V_3  =  h(t)\frac{\partial}{\partial t}
-\rho\dot{h}(t)\frac{\partial}{\partial \rho},\\[3mm]
\ds  V_4  =  l(t)\frac{\partial}{\partial \rho},\qquad
V_5  =  N(t)\frac{\partial}{\partial \theta},
\ea
\]
where $f$, $g$, $h$ and $l$, $N$ are arbitrary functions of $x$, $y$ and $t$
respectively. The nonzero commutation relations between the vector f\/ields are
\[
\ba{l}
\ds \left[V_1(f_1),V_1(f_2)\right]  =  V_1(f_1f'_2-f_2f'_1),\quad
\left[V_2(g_1),V_2(g_2)\right]  =  V_2(g_1g'_2-g_2g'_1),\\[2mm]
\ds \left[V_3(h_1),V_3(h_2)\right]  =  V_3(h_1\dot{h}_2-h_2\dot{h}_1),\quad
\left[V_3(h),V_4(l)\right]  =  V_4(l\dot{h}+h\dot{l}),\\[2mm]
\left[V_3(h),V_5(m)\right]  =  V_5(h\dot{m}).
\ea
\]
All other commutators vanish.

By restricting the arbitrary functions $ f(x)$, $g(y)$, $h(t)$,
$l(t)$ and $N(t)$ to be polynomials in the variables $x$, $y$ and $t$
one can get immediately Virasoro type subalgebras of the form
\[\hspace*{-5.3pt}
\ba{l}
\ds \left[V_1(x^n), V_1(x^m)\right]  =  (m-n)V_1(x^{n+m-1}),\quad
\left[V_2(y^n), V_2(y^m)\right]  =  \frac{-(m-n)}{2A}V_4(y^{n+m-1}),
\\[3mm]
\ds  \left[V_3(t^n), V_3(t^m)\right]  =  -(m-n)2BV_4(t^{n+m-1}),\quad
\left[V_3(t^n), V_4(t^m)\right]  =  (m+n)V_4(t^{n+m-1}),
\\[3mm]
\ds \left[V_3(t^n), V_5(t^m)\right]  =  (m+n)V_5(t^{n+m-1}).
\ea
\]

\renewcommand{\theequation}{D.\arabic{equation}}
\setcounter{equation}{0}

\section{(2+1)-dimensional long dispersive wave equation}

Recently Chakravarthy, Kent and Newman~[33] have introduced a
(2+1)-dimensional long dispersive wave equation of the form
\be \label{lakshmanan:D.1}
\ba{l}
\ds  \lambda q_t+q_{xx}-2q\int (qr)_x d\eta = 0, \\[3mm]
\ds  \lambda r_t-r_{xx}+2r\int (qr)_x d\eta = 0.
\ea
\ee

Eq.(\ref{lakshmanan:D.1}) is the (2+1)-dimensional generalization of the one dimensional
long dispersive wave equation~[34].  By introducing the transformation
$(qr)_x = v_{\eta}$ and rewriting the above equation we get
\be \label{lakshmanan:D.2}
\ba{l}
q_t+q_{xx}-2qv = 0,\\
r_t-r_{xx}+2rv = 0,\\
v_y-rq_x-qr_x = 0.
\ea
\ee

Eq.(\ref{lakshmanan:D.2}) admits P-property and line solitons and dromions~[35].
Eq.(\ref{lakshmanan:D.2}) admits the following vector f\/ields~[18]
\[
V=V_1(f)+V_2(g)+V_3(m)+V(N),
\]
where
\[
\ba{l}
\ds V_1(f) = \frac{x}{2}\dot{f}(t)\frac{\partial}{\partial x}
+f(t)\frac{\partial}{\partial t}
+\left(\frac{1}{8}\ddot{f}(t)qx^2
-\frac{1}{2}q\dot{f}(t)\right)\frac{\partial}{\partial q}\\[3mm]
\ds \qquad  -\frac{1}{8}\ddot{f}(t)x^2 r\frac{\partial}
{\partial r}
+\left(\frac{1}{16}\frac{d^3f}{dt^3}x^2 -v\dot{f}(t)\right)\frac{\partial}
{\partial v},
\ea
\]
\[
V_2(g) = g(t)\frac{\partial}{\partial
x}+\frac{1}{2}\dot{g}(t)xq\frac{\partial}
{\partial q}
-\frac{1}{2}\dot{g}(t)xr\frac{\partial}{\partial r}+
\frac{1}{4}\ddot{g}(t)x\frac{\partial}{\partial v},
\]
\[
V_3(m)=m(y)\frac{\partial}{\partial y}-m'(y)q\frac{\partial}{\partial q},
\qquad
V_4(N)=-qN(y,t)\frac{\partial}{\partial
q}+rN(y,t)\frac{\partial}{\partial r}.
\]

The associated Lie algebra between these vector f\/ields become
\[
\left[V_1(f_1), V_1(f_2)\right] = V_1(f_1\dot{f}_2-f_2\dot{f}_1),
\]
\[
\left[V_2(g_1), V_2(g_2)\right] =
\frac{g_1\dot{g}_2-g_2\dot{g}_1}{2}\left(q\frac{\partial}{\partial q}
-r\frac{\partial}{\partial r}\right)+
\frac{g_1\ddot{g}_2-g_2\ddot{g}_1}{4}\frac{\partial}{\partial v},
\]
\[
\left[V_3(m_1), V_3(m_2)\right] =V_3(m_1m_2'-m_2m_1'),\quad
\left[V_1(f), V_2(g)\right] =
V_2\left(f\dot{g}-\frac{1}{2}g\dot{f}\right),
\]
\[
\left[V_1(f), V_4(N)\right] =V_4(f\dot{N}),\qquad
\left[V_3(m), V_4(N)\right] =V_4(mN'),
\]
which is obviously an inf\/inite dimensional Lie algebra of symmetries.  A
Virasoro-Kac-Moody type subalgebra is immediately obtained by restricting
the arbitrary functions $f$ and $m$ to Laurent polynomials so that we have
the commutators
\[
\left[V_1(t^n), V_1(t^m)\right] =(m-n)V_1(t^{n+m-1}),\qquad
\left[V_3(y^n), V_3(y^m)\right] =(m-n)V_3(y^{n+m-1}).
\]

\label{lakshmanan-lp}

\end{document}